\documentclass[epj]{svjour}
\usepackage{graphicx}
\usepackage{amssymb,amsbsy,amsmath}

\newcommand{\op}[1]{%
    \fontdimen12\textfont3=2pt\fontdimen12\scriptfont3=1.4pt%
    \!\null\mathop{\vphantom{#1}\smash{#1}}\limits_{\sim}\null\!}
\newcommand{\xref}[1]{\protect\ref{#1}}
\newcommand{\figref}[1]{Fig.~\protect\ref{#1}}
\newcommand{\fmref}[1]{(\protect\ref{#1})}
\def\bra#1{\langle \, {#1} \, | \,}
\def\ket#1{\, | \, {#1} \, \rangle}
\newcommand{\braket}[2]{\langle \, {#1} \, | \, {#2} \, \rangle}
\newcommand{\Mean}[1]{\big\langle\big\langle \; {#1}\; 
            \big\rangle\big\rangle}
\newcommand{\dimhm}[2]{D({#1},{#2})}
\newcommand{\MMax}{M_{\text{Max}}}
\newcommand{\pp}[2]{\frac{\partial \, {#1}}{\partial \, {#2}}}
\journalname{Eur. Phys. J. B}
\begin{document}
\title{Application of the finite-temperature Lanczos method for
  the evaluation of magnetocaloric properties of large magnetic molecules}
\titlerunning{Magnetocaloric properties of large magnetic molecules}
\author{J\"urgen Schnack%
 \and Christian Heesing%
}                     
\offprints{J\"urgen Schnack}          
\institute{Department of Physics, Bielefeld University, P.O. box 100131, D-33501 Bielefeld, Germany}
\date{Received: date / Revised version: date}
%
\abstract{
We discuss the magnetocaloric properties of gadolinium
containing magnetic molecules which potentially could be used
for sub-Kelvin cooling. We show that a degeneracy of a singlet
ground state could be advantageous in order to support
adiabatic processes to low temperatures and simultaneously
minimize disturbing dipolar interactions. Since the Hilbert
spaces of such spin systems assume very large dimensions we
evaluate the necessary thermodynamic observables by means of the
Finite-Temperature Lanczos Method.
} 
\maketitle
%

\section{Introduction}
\label{sec-1}

The use of magnetic molecules for sub-Kelvin cooling is one of
the latter ideas in the field of molecular magnetism. Originally
paramagnetic salts have been employed for this purpose which in
1949 led
to the Nobel prize for William Francis Giauque \cite{GiM:PR33}. In recent years
several molecular systems have been discussed with respect to their
magnetocaloric properties \cite{ECG:APL05,MJP:ACIE07,evangelisti:104414,SZT:CC11,ZEW:CS11,HSP:ACIE12,Ses:ANIE12}.
Some aspects have turned out to be of importance if one wants to
compete with the cooling power of paramagnetic salts: the
low-lying energy eigenstates should 
possess large magnetic moments and they should be dense, so that
one can sweep many of them with moderate magnetic fields. The magnetic
anisotropy should be small, otherwise energy barriers could
prevent a high density of states. These aspects make gadolinium
the preferred metal ion: it has a large spin quantum number of
$s=7/2$, a rather small exchange interaction in chemical
complexes and a negligible single-ion anisotropy, compare
e.g. Refs.~\cite{SZT:CC11,ZEW:CS11,HSP:ACIE12}. 

From the point of view of theoretical modeling gadolinium
compounds are demanding since the sizes of the respective Hilbert
spaces grow very rapidly with the number of involved gadolinium
ions due to the large spin quantum number. Even when employing
all possible symmetries the system size is rather restricted
since the dimension of the largest subspace should not exceed a
size of 100,000 in order to render a complete numerical diagonalization
possible \cite{ScS:PRB09,ScS:IRPC10}. Fortunately, a very
accurate approximation has been developed for cases with
Hilbert space dimensions of up to roughly $10^{10}$ -- the
Finite Temperature Lanczos Method
(FTLM) \cite{PhysRevB.49.5065,JaP:AP00}. In a recent publication
we demonstrated that this method is indeed capable of evaluating
thermodynamic observables for magnetic molecules with an accuracy
that is nearly indistinguishable from exact
results \cite{ScW:EPJB10}.

The interesting physical question is, which arrangements of
interacting magnetic ions permit a large magnetocaloric effect. For this
purpose the low-lying zero-field density of states should be
high, and an 
applied field should fan out these levels. Magnetic frustration
is known to yield dense spectra and to enhance the
magnetocaloric effect under certain
circumstances \cite{Zhi:PRB03,SSR:PRB07,HoZ:JPCS09}. 
In addition frustration, i.e. a situation where not all pairs
of spins assume a collinear arrangement in the classical ground
state \cite{Sch:DT10}, may result in
singlet ground states which are degenerate. This would have two
advantageous consequences: the ground state would be connected
to an isentrope with $S>0$ which should allow for very
low-temperature cooling. In addition a singlet ground state
minimizes the dipolar interactions compared to high-spin
molecules where dipolar interactions prevent a cooling to very
low-temperatures \cite{MME:AM12}. We investigate our 
hypothesis with the simplest of such systems, the tetrahedron.

It should be mentioned that one can also aim at designing the
intermolecular interactions in order to reduce the influence of
dipolar interactions \cite{MME:AM12}. The most simple way
consists in increasing the mutual distances by means of bulky
ligands. This route will not be discussed in the present
article. 

The article is organized as follows. In Section~\xref{sec-2}
basics of the finite-temperature Lanczos method are
repeated. Section~\xref{sec-3} is devoted to the discussion of
several recently synthesized gadolinium containing compounds.
The article closes with a summary. A technical appendix explains
a basis coding scheme used in the FTLM.

\section{The finite-temperature Lanczos method}
\label{sec-2}

For the evaluation of thermodynamic properties in the canonical
ensemble the exact partition function $Z$ depending on
temperature $T$ and magnetic field $B$ is given by 
\begin{eqnarray}
\label{E-1-1}
Z(T,B)
&=&
\sum_{\nu}\;
\bra{\nu} e^{-\beta \op{H}} \ket{\nu}
\ .
\end{eqnarray}
Here $\{\ket{\nu}\}$ denotes an orthonormal basis of the
respective Hilbert space. Following the ideas of
Refs.~\cite{PhysRevB.49.5065,JaP:AP00} the unknown matrix
elements are approximated as
\begin{eqnarray}
\label{E-1-2}
\bra{\nu} e^{-\beta \op{H}} \ket{\nu}
&\approx&
\sum_{n=1}^{N_L}\;
\braket{\nu}{n(\nu)} e^{-\beta \epsilon_n^{(\nu)}} \braket{n(\nu)}{\nu}
\ .
\end{eqnarray}
For the evaluation of the right hand side of Eq.~\fmref{E-1-2}
$\ket{\nu}$ is taken as the initial vector of a Lanczos
iteration. This iteration consists of $N_L$ Lanczos steps, which
span a respective Krylow space. As common for the Lanczos method
the Hamiltonian is diagonalized in this Krylow space. This
yields the $N_L$ Lanczos eigenvectors $\ket{n(\nu)}$ as well as
the associated Lanczos energy eigenvalues
$\epsilon_n^{(\nu)}$. They are enumerated by $n=1,\dots,
N_L$. The notation $n(\nu)$ is chosen to remind one that the
Lanczos eigenvectors $\ket{n(\nu)}$ belong to the Krylow space
derived from the original state $\ket{\nu}$.

The number of Lanczos steps $N_L$ is a parameter of the
approximation that needs to be large enough to reach the
extremal energy eigenvalues but should not be too large in order
not to run into problems of numerical accuracy. $N_L\approx 100$
is a typical and good value.

In addition, the complete and thus very large sum over all
states $\ket{\nu}$ is 
replaced by a summation over a subset of $R$ random
vectors. These vectors are truly random, they do not need to
belong to any special basis set. Altogether this yields for the
partition function
\begin{eqnarray}
\label{E-1-3}
Z(T,B)
&\approx&
\frac{\text{dim}({\mathcal H})}{R}
\sum_{\nu=1}^R\;
\sum_{n=1}^{N_L}\;
e^{-\beta \epsilon_n^{(\nu)}} |\braket{n(\nu)}{\nu}|^2
\ .
\end{eqnarray}
It will of courseimprove the accuracy if symmetries are taken
into account as in the following formulation
\begin{eqnarray}
\label{E-1-4}
Z(T,B)
&\approx&
\sum_{\Gamma}\;
\frac{\text{dim}({\mathcal H}(\Gamma))}{R_{\Gamma}}
\sum_{\nu=1}^{R_{\Gamma}}\;
\sum_{n=1}^{N_L}\;
\nonumber \\
&&\times
e^{-\beta \epsilon_n^{(\nu,\Gamma)}} |\braket{n(\nu, \Gamma)}{\nu, \Gamma}|^2
\ .
\end{eqnarray}
$\Gamma$ labels the irreducible representations of the
employed symmetry group. The complete Hilbert space is decomposed
into mutually orthogonal subspaces ${\mathcal H}(\Gamma)$.

An observable would then be calculated as
\begin{eqnarray}
\label{E-1-5}
O(T,B)
&\approx&
\frac{1}{Z(T,B)}
\sum_{\Gamma}\;
\frac{\text{dim}({\mathcal H}(\Gamma))}{R_{\Gamma}}
\sum_{\nu=1}^{R_{\Gamma}}\;
\sum_{n=1}^{N_L}\;
e^{-\beta \epsilon_n^{(\nu,\Gamma)}}
\nonumber \\
&&\times
\bra{n(\nu, \Gamma)}\op{O}\ket{\nu, \Gamma}
\braket{\nu, \Gamma}{n(\nu, \Gamma)}
\ .
\end{eqnarray}
This approximation of the observable $O(T,B)$ may contain large 
statistical fluctuations at low temperatures due to the
randomness of the set of states $\{\ket{\nu, \Gamma}\}$, but
this can be cured by assuming a symmetrized version of
Eq.~\fmref{E-1-5} \cite{PhysRevB.67.161103}. For our
investigations this is irrelevant.

In this article the entropy plays a central role, it is
evaluated as 
\begin{eqnarray}
\label{E-1-6}
S(T,B)
&=&
\Mean{\op{H}}/T + k_B \log(Z(T,B)) 
\ ,
\end{eqnarray}
with $Z(T,B)$ being calculated according to \eqref{E-1-4} and 
$\Mean{\op{H}}$ according to \eqref{E-1-5}.

Our very positive experience is that even for large problems the
number of random starting vectors as well as the number of
Lanczos steps can be chosen rather small, e.g. $R\approx 20,
N_L\approx 100$, compare Ref.~\cite{ScW:EPJB10}.

\section{Magnetocalorics of certain gadolinium compounds}
\label{sec-3}

The following spin systems are described by the Heisenberg spin
Hamiltonian augmented with a Zeeman term, i.e.
\begin{eqnarray}
\label{E-3-1}
\op{H}
&=&
-
2\;
\sum_{i<j}\;
{J}_{ij}
\op{\vec{s}}_i \cdot \op{\vec{s}}_j
+
g\, \mu_B\, B\,
\sum_{i}\;
\op{s}^z_i
\ .
\end{eqnarray}
${J}_{ij}$ is the exchange parameter between spins at sites $i$
and $j$. For the sake of simplicity it is assumed that all spins
have the same $g$-factor. Since $\left[\op{H}, \op{{S}}^z\right]
= 0$, this (simple) 
symmetry is used for the finite-temperature Lanczos
calculations.

In a process of adiabatic demagnetization the temperature
changes with field according to the following thermodynamic
relation: 
\begin{eqnarray}
\label{E-3-2}
\left(\pp{T}{B}\right)_S
&=&
-\frac{T}{C}\left(\pp{S}{B}\right)_T
\ .
\end{eqnarray}
Here $S$ denotes entropy, $T$ temperature, $B$ magnetic
induction, and $C$ heat capacity. Besides the heat capacity,
which in our examples does not vary too much for small fields
and temperatures $T\approx 1$~K, the isothermal entropy
change $\left(\pp{S}{B}\right)_T$ has a large
impact \cite{Ses:ANIE12}. At very low temperatures close to $T=0$
the isothermal entropy change is of course very large if the ground state of the
magnetic molecule possesses a large total spin quantum number $S_t$, since
this corresponds to a theoretical entropy of $S(T=0,B=0)=k_B
\log(2S_t+1)$.\footnote{Here a short remark concerning the third law of
thermodynamics might be necessary: The third law conjectures
that at $T=0$ the entropy of any system is a universal constant
that can be taken to be $S=0$ which in turn means that the
ground state is non-degenerate. Models, however, can show a
ground state degeneracy and thus a residual entropy at $T=0$. In
reality tiny interactions might split this degeneracy at their
energy scale, which means that the residual entropy remains at
its value down to this scale.}
Therefore, naively thinking, all isentropes with an
entropy equal to or smaller than this value should run into
absolute zero. How close they come in reality depends on the
very small interactions that become relevant at very low
temperatures. The most disturbing interaction for high-spin
molecules is the dipolar interaction, which also in the
case of paramagnetic salts limits the achievable temperatures. In
the following we therefore also discuss a possible way out of
this dilemma: molecules that possess an $S_t=0$ ground state with
residual entropy due to a ground state degeneracy.

\subsection{Gd$_4$Cu$_8$ \& Gd$_4$Ni$_8$}
\label{sec-3-1}

The M=Cu and M=Ni members of the family of Gd$_4$M$_8$ molecules
were synthesized quite recently \cite{HSP:ACIE12}. The
eigenvalues of the respective spin Hamiltonians could be
determined numerically exactly for the case of Gd$_4$Cu$_8$, but
not for Gd$_4$Ni$_8$. In the latter case the Finite-Temperature
Lanczos Method was employed.

For Gd$_4$Cu$_8$ the model Hamiltonian \fmref{E-3-1} includes
the following parameters: 
$J_{\text{Gd}\text{Gd}}=-0.1$~cm$^{-1}$,\linebreak
$J_{\text{Gd}\text{Cu}}=+0.9$~cm$^{-1}$,
$J_{\text{Cu}\text{Cu}}=-8.0$~cm$^{-1}$. The spectroscopic
splitting factor was taken as $g=2.0$.
For Gd$_4$Ni$_8$ the model parameters were chosen as: 
$J_{\text{Gd}\text{Gd}}=-0.1$~cm$^{-1}$,
$J_{\text{Gd}\text{Ni}}=+0.17$~cm$^{-1}$,
$J_{\text{Ni}\text{Ni}}=+12.0$~cm$^{-1}$. Again we took $g=2.0$.
In the following all other interactions or
corrections such as temperature independent paramagnetism,
different $g$ factors for different ions or a possible single-ion
anisotropy in the case of nickel have been neglected.
Despite these approximations all theoretical curves agree nicely
with the experimental ones published in
Ref.~\cite{HSP:ACIE12}.

\begin{figure}[ht!]
\centering
\includegraphics*[clip,width=75mm]{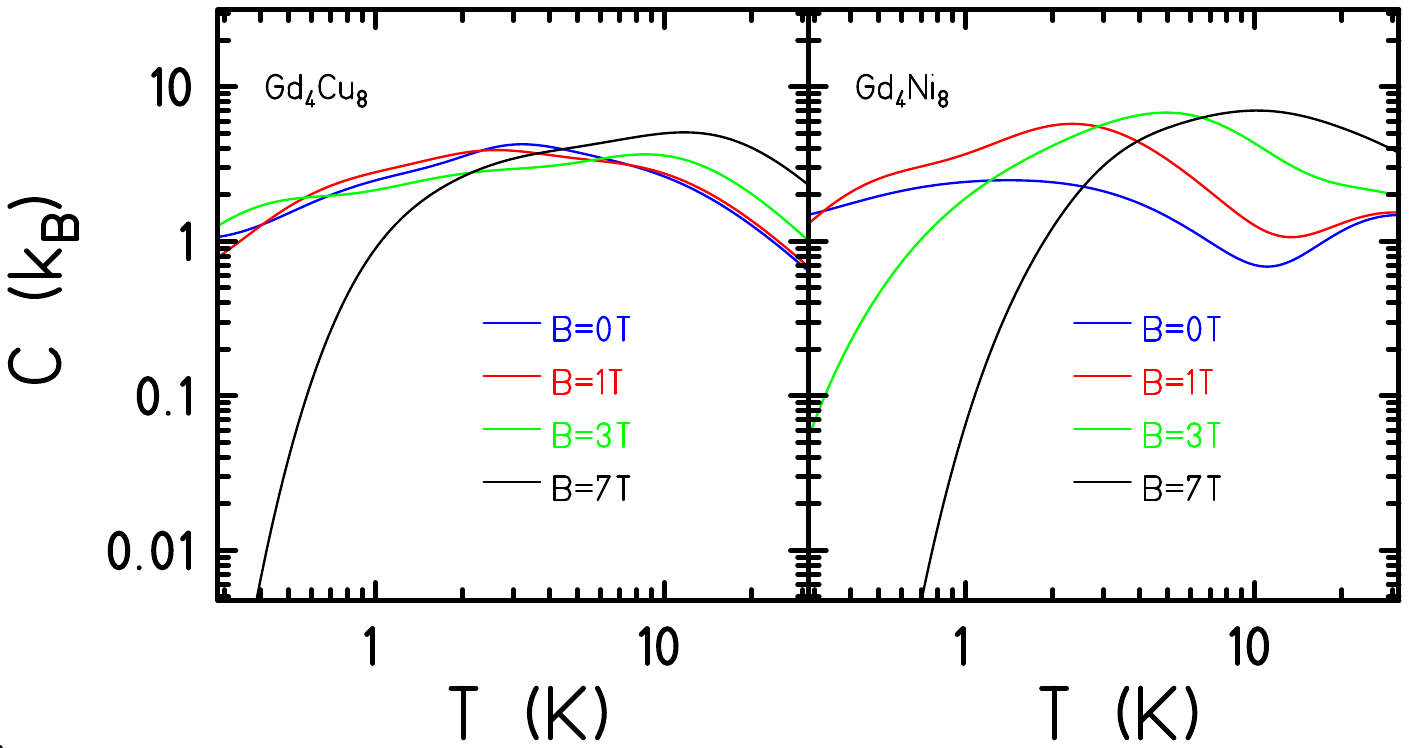}
\caption{Theoretical heat capacity per molecule for Gd$_4$Cu$_8$
  (l.h.s.) and Gd$_4$Ni$_8$ (r.h.s.)
  at various magnetic fields.}
\label{ftlmmce-f-1}
\end{figure}

Figure~\xref{ftlmmce-f-1} displays the theoretical heat capacity
per 
\linebreak
molecule for Gd$_4$Cu$_8$ (l.h.s.) and Gd$_4$Ni$_8$ (r.h.s.)
at various magnetic fields. The behavior is for both compounds
qualitatively similar.

\begin{figure}[ht!]
\centering
\includegraphics*[clip,width=75mm]{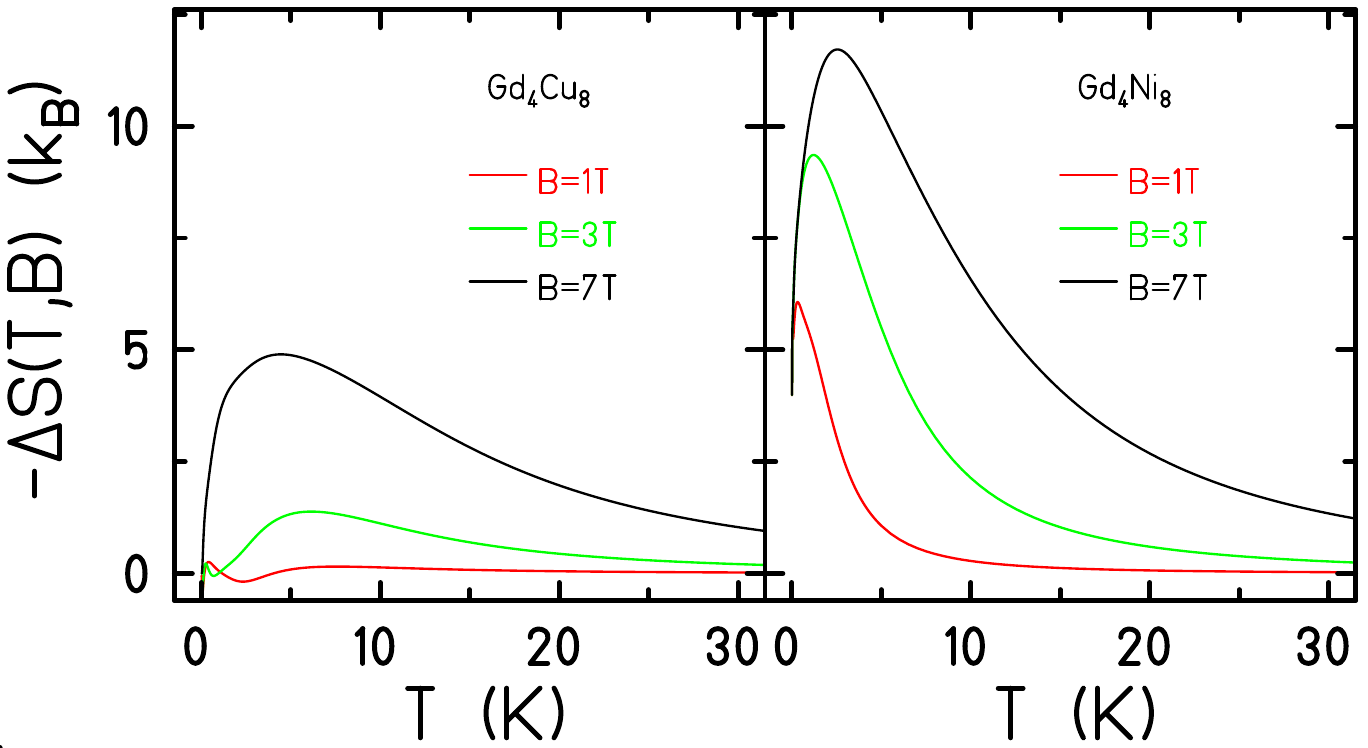}
\caption{Theoretical isothermal entropy change per molecule for
   Gd$_4$Cu$_8$ (l.h.s.) and Gd$_4$Ni$_8$ (r.h.s.) 
   for various field differences:$-\Delta
   S(T,B)=-[S(T,B)-S(T,0)]$.}
\label{ftlmmce-f-2}
\end{figure}

The isothermal magnetic entropy change, compare
\linebreak
\figref{ftlmmce-f-2}, 
turns out to be very 
different; it is much larger for Gd$_4$Ni$_8$. The reason is
that for Gd$_4$Ni$_8$ the low-lying multiplets belong to large
total spin quantum 
numbers which leads to larger entropies at low
temperatures. This is made even clearer in the two following
plots displaying the isentropes as function of both temperature
and magnetic field.

\begin{figure}[ht!]
\centering
\includegraphics*[clip,width=75mm]{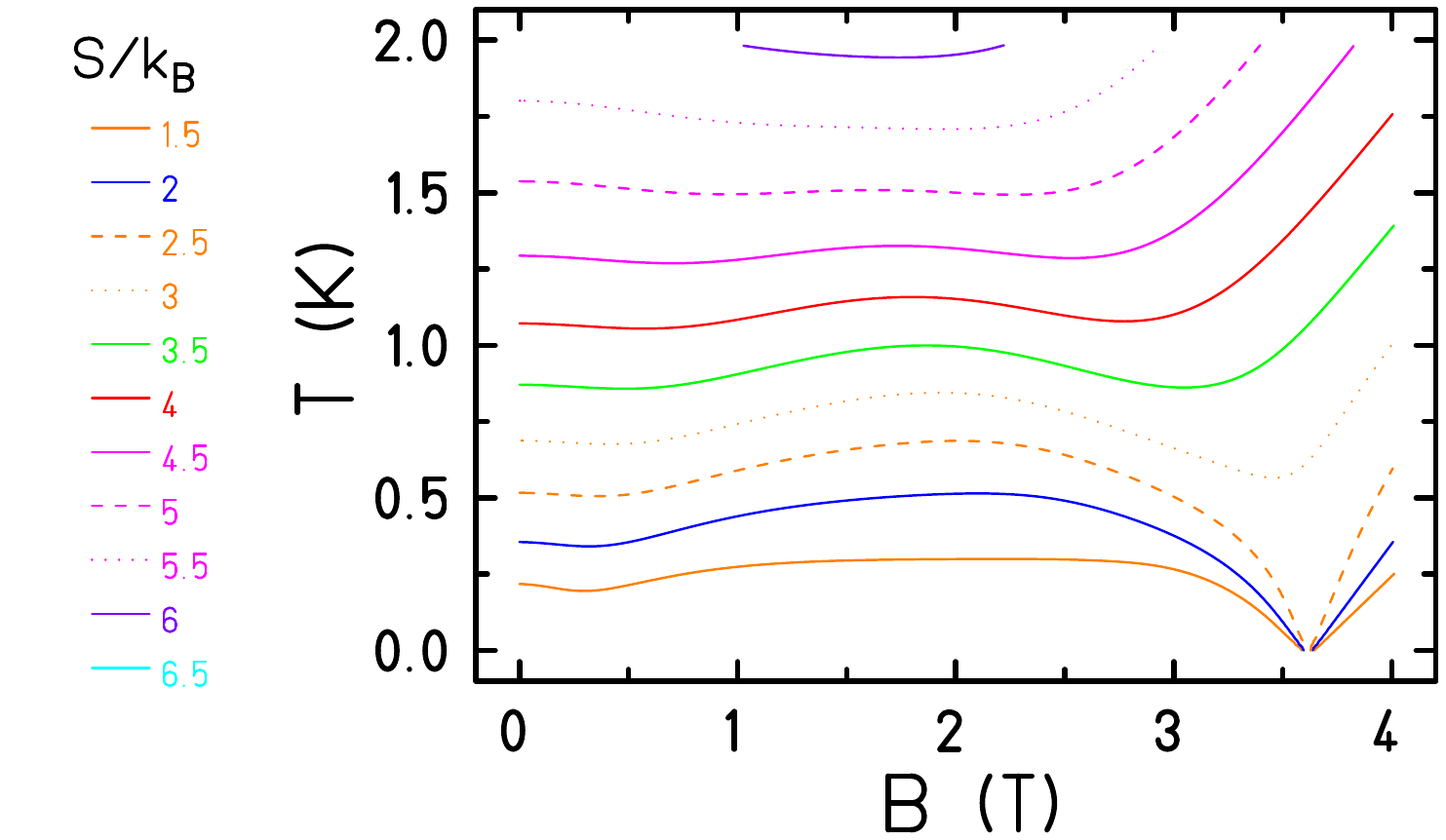}
\caption{Theoretical isentropes for Gd$_4$Cu$_8$.} 
\label{ftlmmce-f-3}
\end{figure}

\begin{figure}[ht!]
\centering
\includegraphics*[clip,width=75mm]{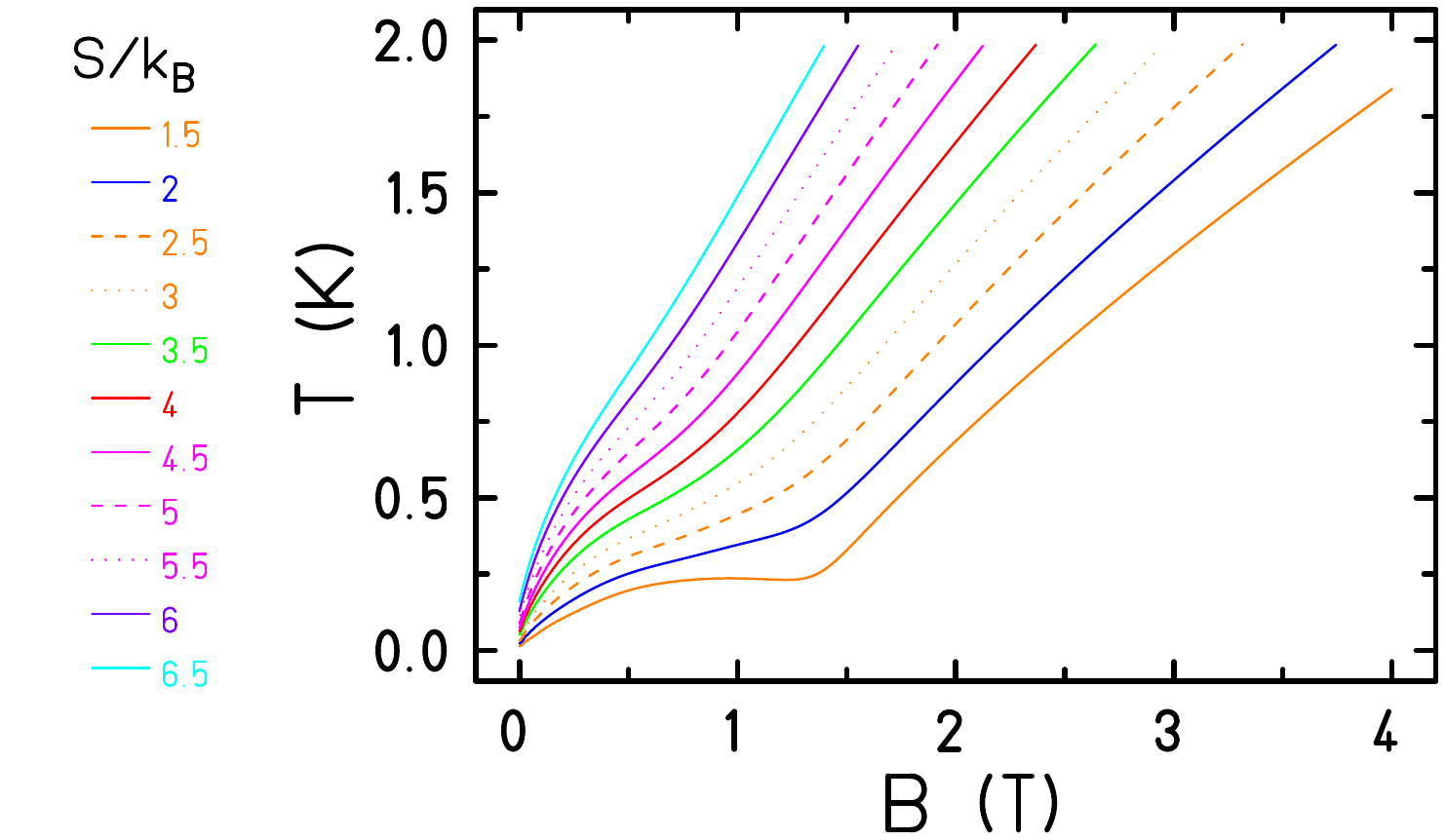}
\caption{Theoretical isentropes for Gd$_4$Ni$_8$.} 
\label{ftlmmce-f-4}
\end{figure}

Gd$_4$Cu$_8$ (\figref{ftlmmce-f-3}) possesses a non-degenerate
$S_t=0$ ground state that is separated from a triplet and a
quintet, whereas Gd$_4$Ni$_8$ (\figref{ftlmmce-f-4}) has a
ground state with $S_t=22$. In the latter case all isentropes with 
$S \leq k_B \log(45)$ run into absolute zero, which is clearly
visible in \figref{ftlmmce-f-4}. On the contrary, since
Gd$_4$Cu$_8$ possesses a non-degenerate $S_t=0$ ground state all
isentropes approach temperatures $T>0$ when $B$ goes to zero.

Although this behavior suggests that Gd$_4$Ni$_8$ should be a
very good refrigerant, this does not need to be the case. At
sub-Kelvin temperatures dipolar interactions become very
important. They prevent a closer approach of $T=0$
\cite{MME:AM12}. Dipolar
interactions could be tamed by molecules that possess an
$S_t=0$ ground state, but a non-degenerate ground state would not
be helpful due to its vanishing entropy. Therefore, we suggest
to investigate molecules which have a degenerate -- the more the
better -- ground state with $S_t=0$. A ground state degeneracy can
be induced by frustration, thus a tetrahedron with
antiferromagnetic coupling would be a first candidate.

\subsection{A fictitious Gd$_4$ tetrahedron}
\label{sec-3-3}

About half a dozen Gd$_4$ tetrahedra have been synthesized to
date, none of them was magnetically characterized
\cite{PBH:CC94,MZG:NJC00,MZG:ANIE00,RoU:DT06,KMK:JACS06,KLZ:IC09}.
In the following we therefore discuss the 
magnetic properties of a fictitious Gd$_4$ tetrahedron with an
exchange interaction of 
$J_{\text{Gd}\text{Gd}}=-0.1$~cm$^{-1}$ and $g=2.0$. The
magnetic heat capacity (not shown) looks pretty similar to those
already shown; again $C\approx 1 k_B$ for low fields at
$T\approx 1$~K. 

\begin{figure}[ht!]
\centering
\includegraphics*[clip,width=55mm]{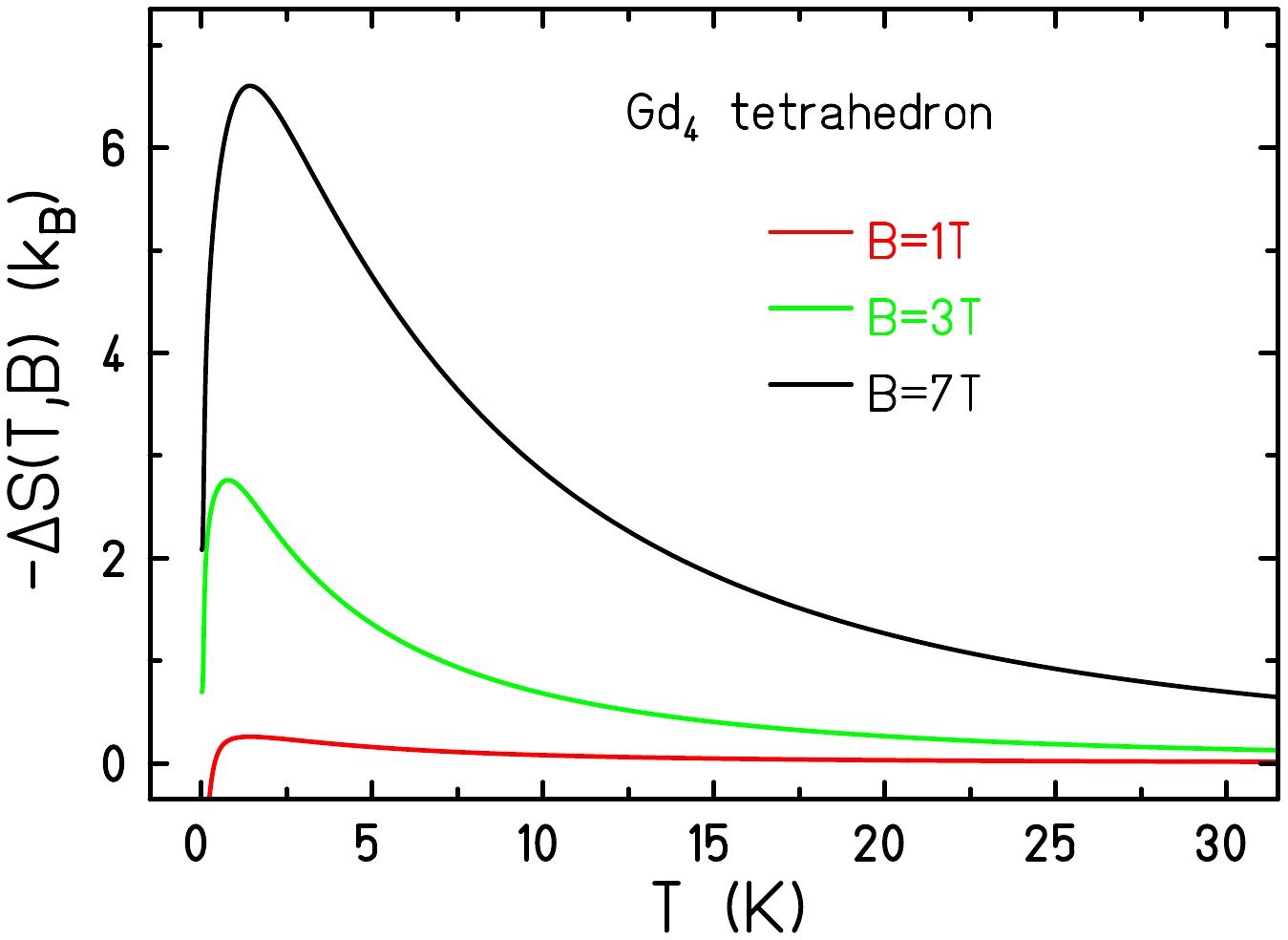}
\caption{Theoretical isothermal entropy change per molecule for
   a Gd$_4$ tetrahedron for various field differences.}
\label{ftlmmce-f-5}
\end{figure}

\begin{figure}[ht!]
\centering
\includegraphics*[clip,width=75mm]{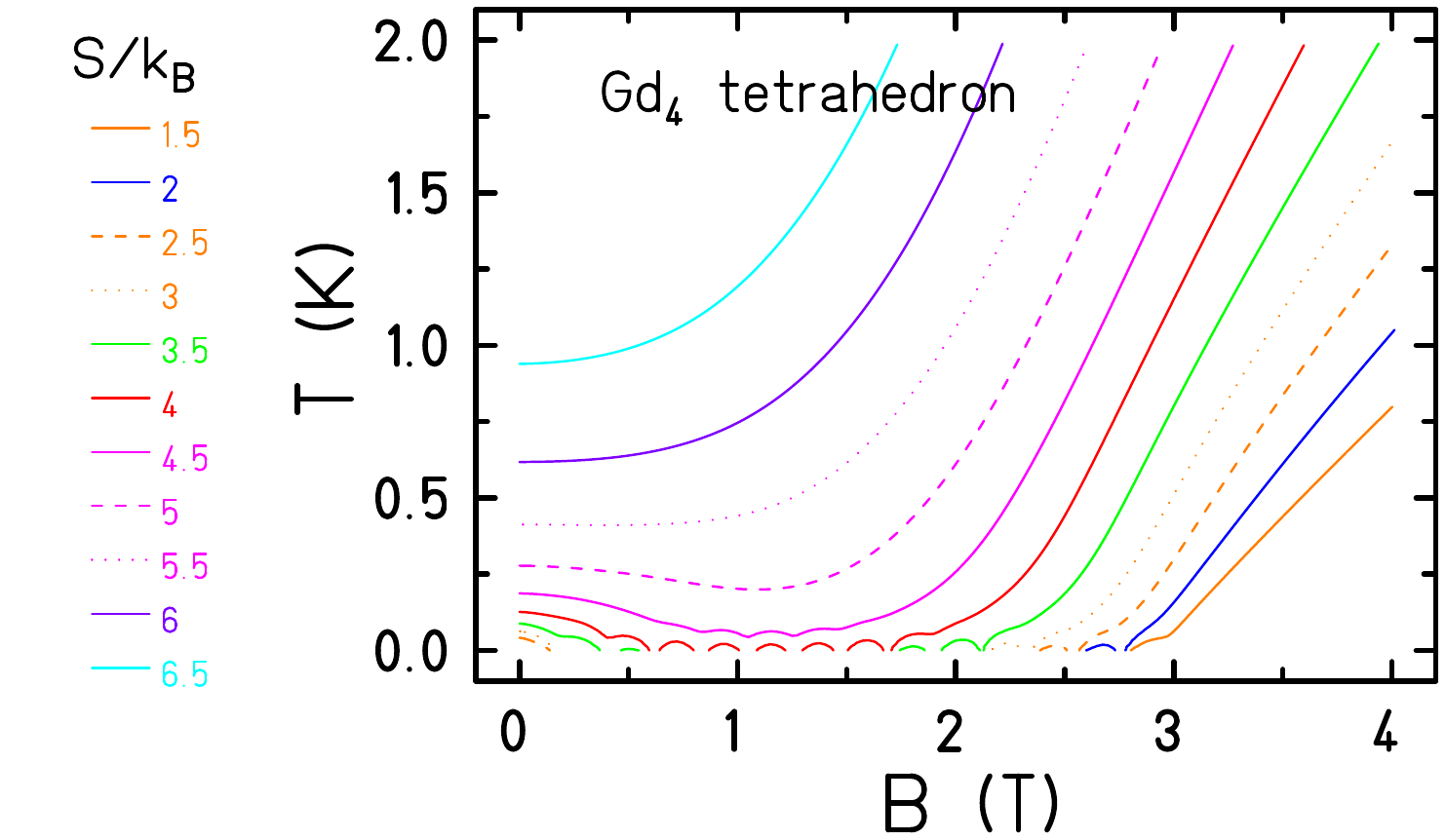}
\caption{Theoretical isentropes for a Gd$_4$ tetrahedron. The
  lowest isentropes overlap at $T=0$ in this linear plot.}
\label{ftlmmce-f-6}
\end{figure}

The isothermal magnetic entropy change, compare
\linebreak
\figref{ftlmmce-f-5}, looks unspectacular. Nevertheless, the
isentropes shown in \figref{ftlmmce-f-8} demonstrate an unusual
behavior which is not obvious at first glance. Since the $S=0$
ground state is eightfold degenerate all isentropes with
$S \leq k_B \log(8)=2.08 k_B$ run into absolute zero. But for a
tetrahedron this happens in a different way compared to
molecules with high-spin ground state such as
Gd$_4$Ni$_8$. The isentropes of high-spin molecules
approach zero like a paramagnet with a rate of 
\begin{eqnarray}
\label{E-3-3}
\left(\pp{T}{B}\right)_S
&\approx &
\frac{T}{B}
\ ,
\end{eqnarray}
i.e. rather steeply. For molecules with a diamagnetic ground
state zero is approached on a rather ``flat" trajectory,
e.g. from high magnetic fields values like $B=4$~T at $T=1$~K,
compare \figref{ftlmmce-f-6}.

\begin{figure}[ht!]
\centering
\includegraphics*[clip,width=75mm]{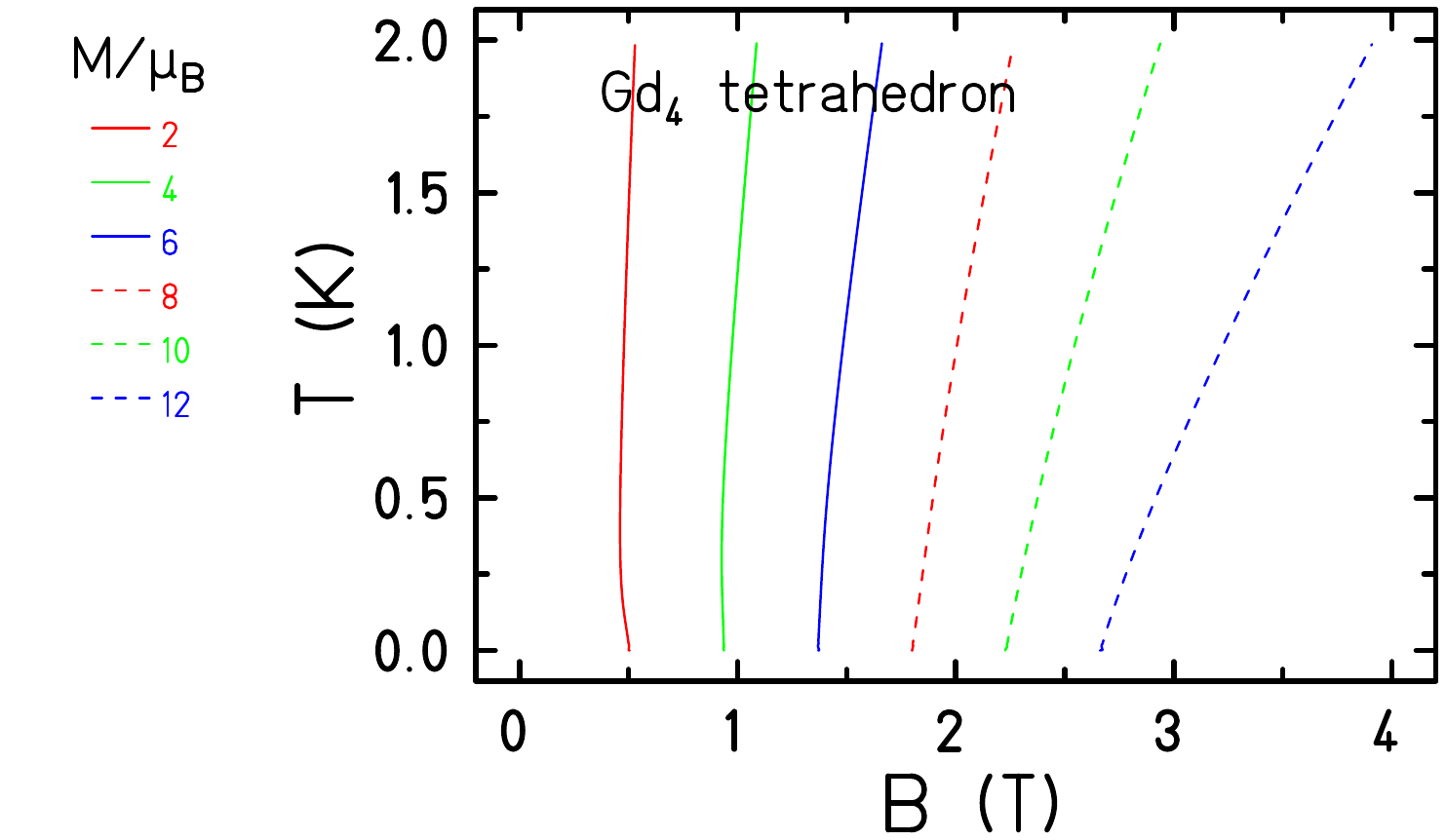}
\caption{Theoretical magnetization contours for a Gd$_4$
  tetrahedron.} 
\label{ftlmmce-f-7}
\end{figure}

Figure \xref{ftlmmce-f-7} shows the related magnetization
contour plot. While decreasing the magnetic field also the
magnetization decreases and consequently also the dipolar
interaction. 

\subsection{A Gd$_6$ octahedron}
\label{sec-3-4}

\begin{figure}[ht!]
\centering
\includegraphics*[clip,width=75mm]{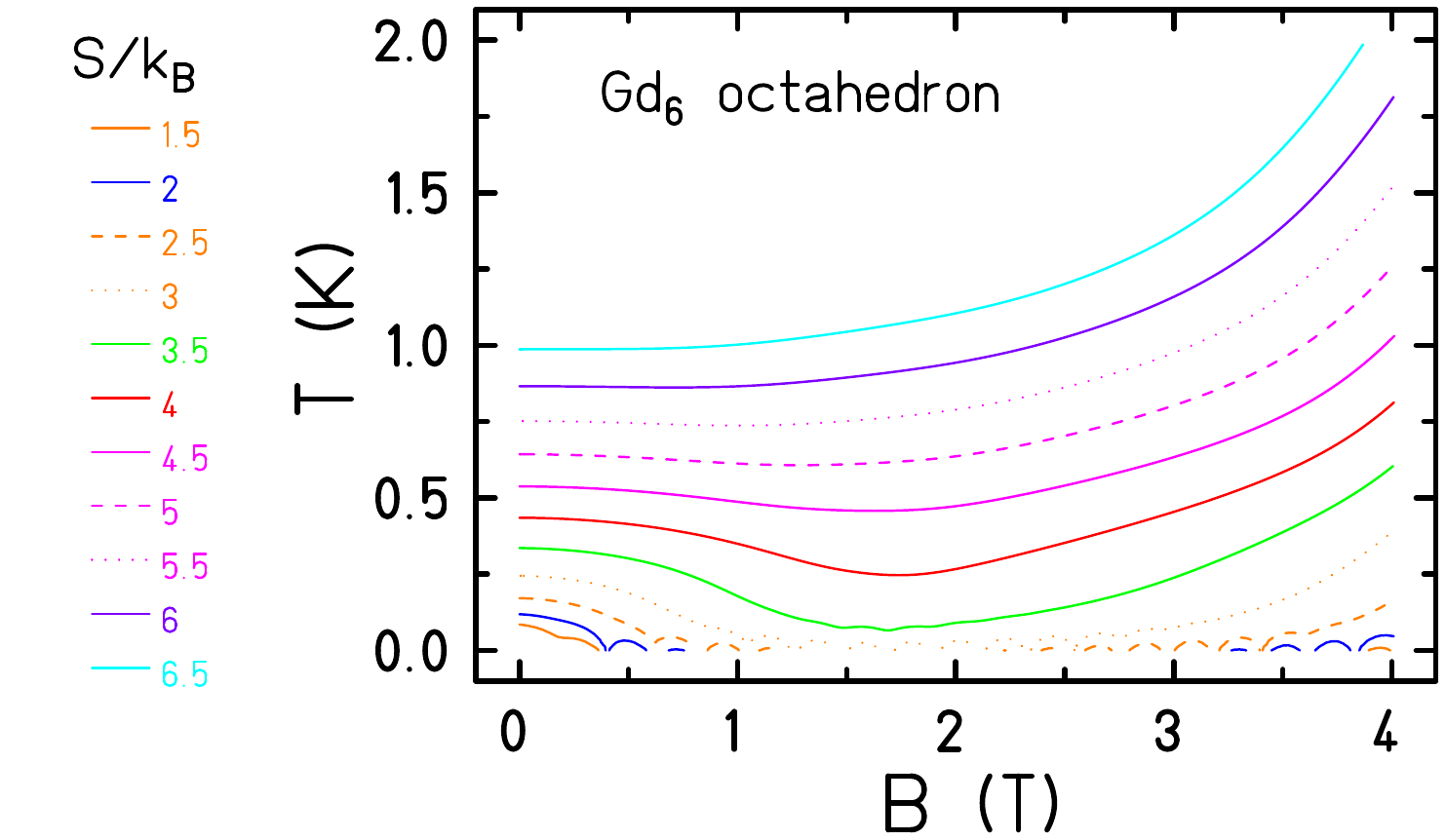}
\caption{Theoretical isentropes for a Gd$_6$ octahedron. The
  lowest isentropes overlap at $T=0$ in this linear plot.}
\label{ftlmmce-f-8}
\end{figure}

As a last example we would like to discuss the octahedron. This
is another interesting structure since it is not too complicated
to be synthesized and it has an interesting spectrum. The
spectrum of an octahedron is that of a so-called
three-sublattice antiferromagnet \cite{SLM:EPL01}. The
non-degenerate ground state possesses $S_t=0$, higher-lying
multiplets are highly degenerate beyond their usual degeneracy
due to magnetic sublevels. The weak point nevertheless is the
non-degenerate ground state which precludes successful cooling,
compare isentropes in \figref{ftlmmce-f-8}.

In addition to the perfect octahedron we would like to discuss a
recently synthesized distorted octahedron \cite{SMB:CC12}. We
assume that this octahedron has an approximate $C_4$
symmetry. Four spins are arranged on the vertices of a square,
one at the top, another one at the bottom. A rough simulation of
the available experimental data yielded the following exchange
integrals: the interaction between top (or bottom) and every
spin of the square 
$J_{\text{t}\text{s}}=J_{\text{b}\text{s}}=-0.05$~cm$^{-1}$,
between nearest neighbors on the square
$J_{\text{s}\text{s}}=-0.02$~cm$^{-1}$, and between top and
bottom spins
$J_{\text{t}\text{b}}=-0.2$~cm$^{-1}$. 
The spectroscopic
splitting factor was taken as $g=2.0$.

\begin{figure}[ht!]
\centering
\includegraphics*[clip,width=60mm]{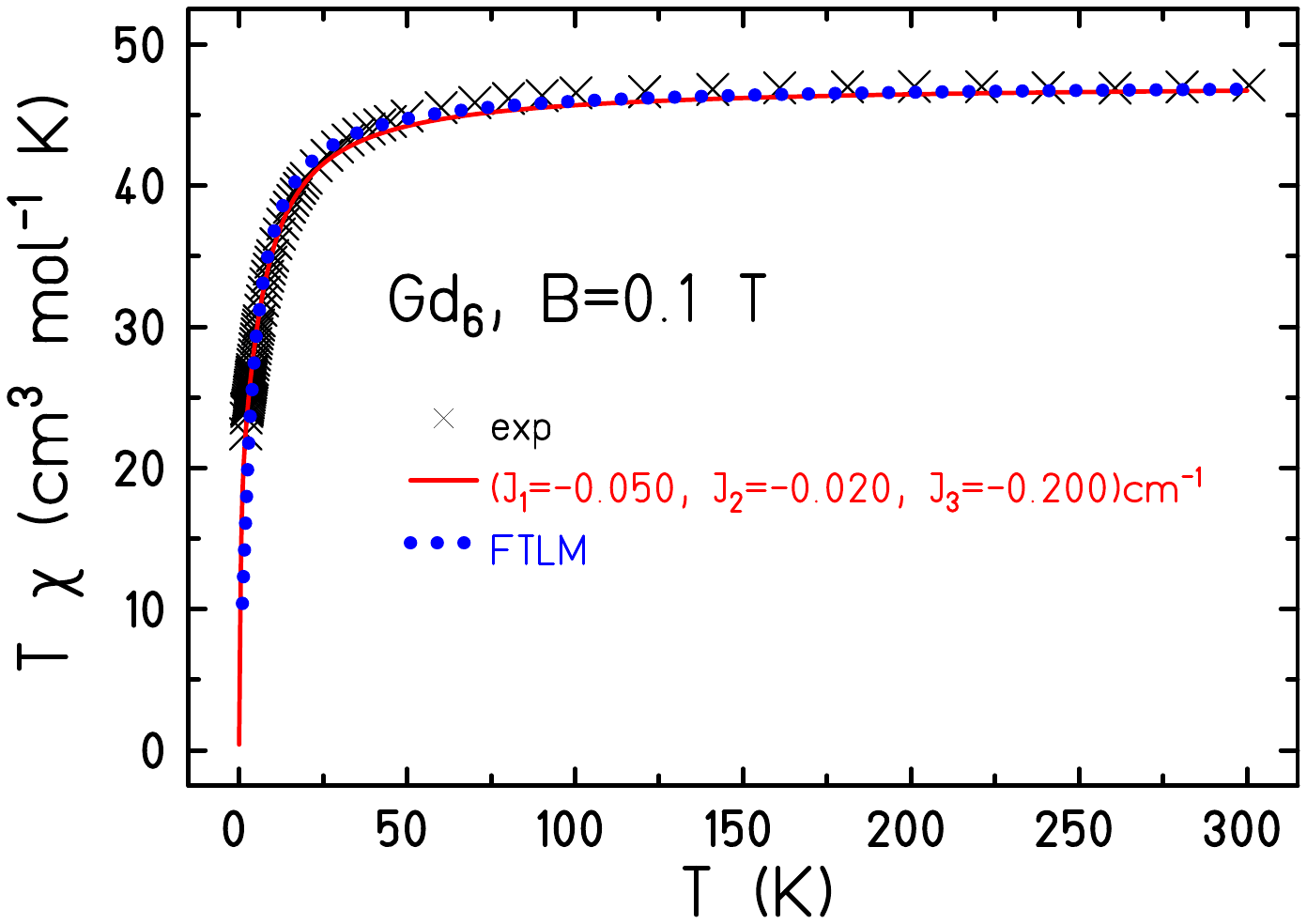}
\caption{Experimental and theoretical susceptibility of the
   distorted octahedron. Experimental values taken from
   Ref.~\cite{SMB:CC12}. The solid curve is the result of a
   complete diagonalization, the dots result from the FTLM.} 
\label{ftlmmce-f-9}
\end{figure}

\begin{figure}[ht!]
\centering
\includegraphics*[clip,width=60mm]{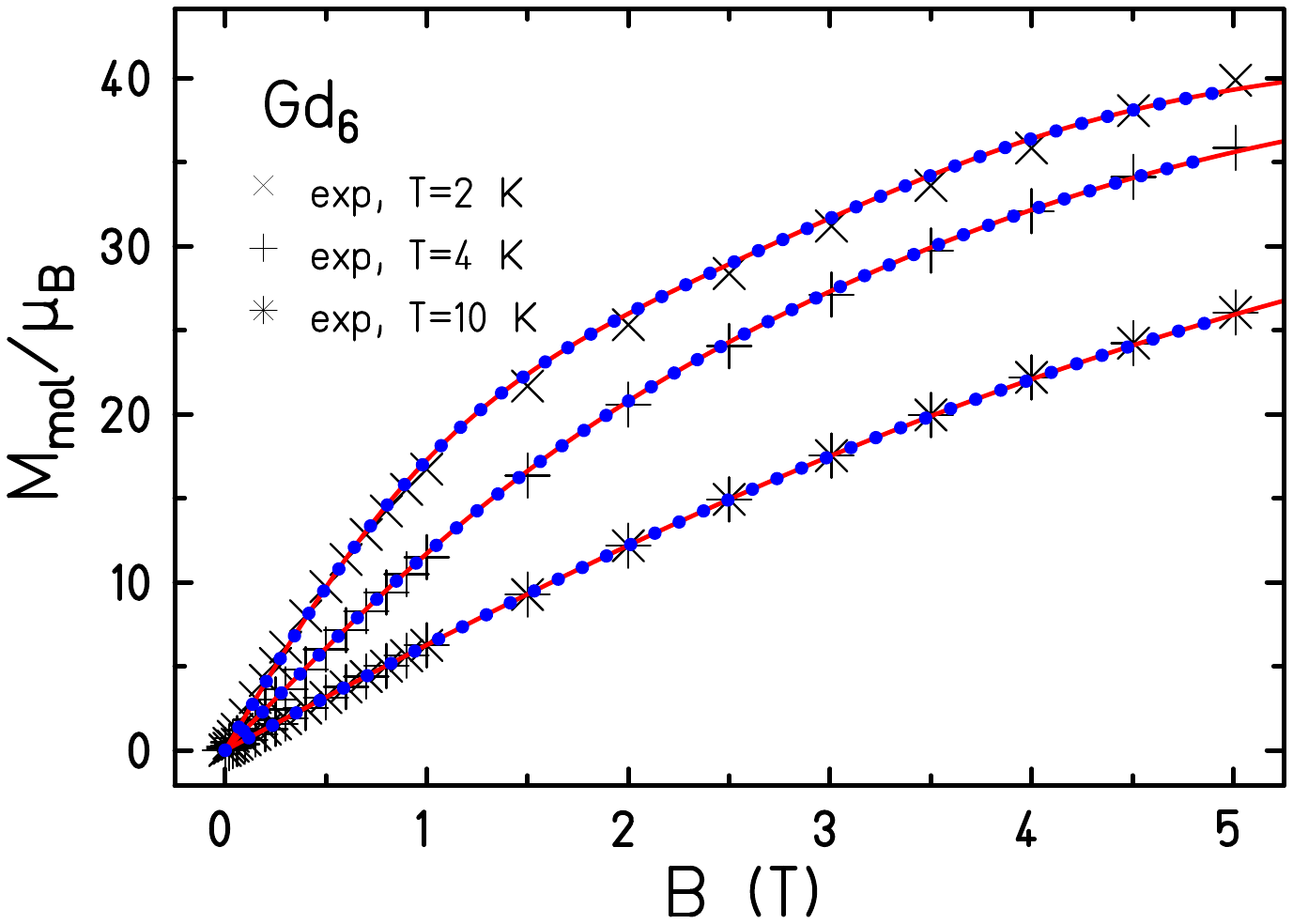}
\caption{Experimental and theoretical magnetization of the
   distorted octahedron. Experimental values taken from
   Ref.~\cite{SMB:CC12}. The solid curve is the result of a
   complete diagonalization, the dots result from the FTLM.} 
\label{ftlmmce-f-10}
\end{figure}

\begin{figure}[ht!]
\centering
\includegraphics*[clip,width=75mm]{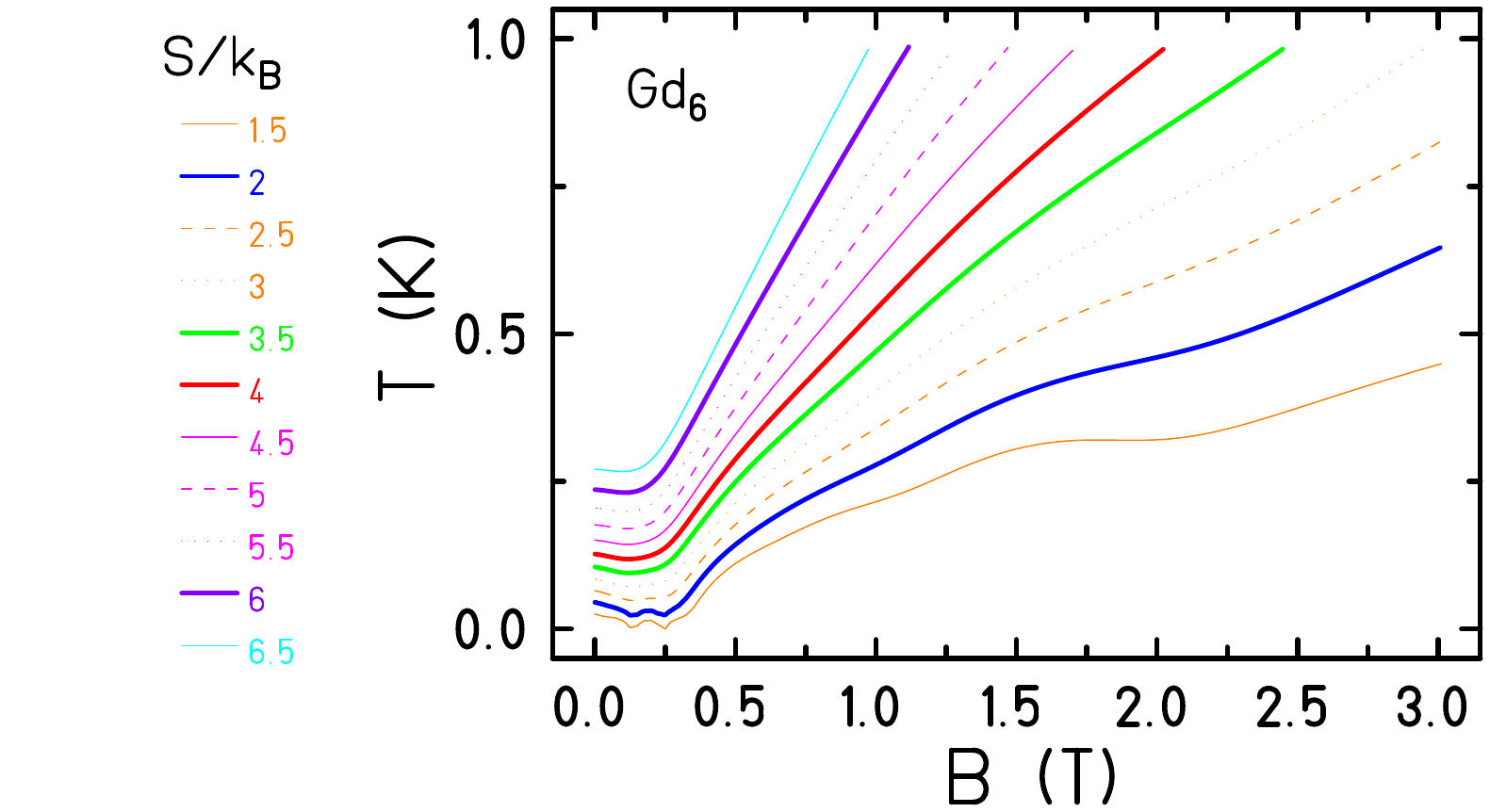}
\caption{Theoretical isentropes for the distorted Gd$_6$ octahedron.}
\label{ftlmmce-f-11}
\end{figure}

As one can see in Figs.~\xref{ftlmmce-f-9} and
\xref{ftlmmce-f-10} a coupling scheme with the above given exchange
interactions yields a very good approximation of the experimental
data. We can now predict how this material would behave as an
adiabatic cooler. Figure~\xref{ftlmmce-f-11} depicts the
isentropes of the distorted octahedron.

Although also this molecule has a non-degenerate diamagnetic
ground state the isentropes exhibit a much steeper slope,
i.e. larger cooling rate, compared to the regular octahedron. We
conjecture that this results from the fact that the
non-symmetric, i.e. only $C_4$ symmetric interactions split the
highly degenerate multiplets and thus lead to a smeared out
density of states. In addition, some of the interactions are
smaller than in the example or a regular octahedron which also
reduces the size of low-lying gaps.

\section{Summary and Outlook}
\label{sec-5}

The Finite-Temperature Lanczos Method enabled us to evaluate the
thermal properties of larger gadolinium containing magnetic
molecules. Due to the large intrinsic spin of gadolinium these
substances are potentially useful sub-Kelvin coolers. A major
quality criterion is the achievable ground state degeneracy. If
such a degeneracy could be realized for a singlet ground state
this would also minimize disturbing dipolar interactions.

\section*{Acknowledgment}

We are deeply indebted to Euan Brechin and Thomas Hooper for
informing us about the synthesized Gd tetrahedra as well as
their distorted Gd$_6$ octahedron, and we would also like to
thank Marco Evangelisti very much for very valuable comments
from an experimental point of view.
This work was supported by the German Science Foundation (DFG)
through the research group 945. Computing time at the Leibniz
Computing Center in Garching is also gratefully
acknowledged. Last but not least we like to thank the State of
North Rhine-Westphalia and the DFG for financing our local SMP
supercomputer as well as the companies BULL and ScaleMP for
their support.

\appendix
\section{Basis coding for mixed spin systems}
\label{sec-a}

Since Lanczos iterations consist of matrix vector
multiplications they can be parallelized by \verb§openMP§
directives \cite{SHS:JCP07}. In our programs this is further accelerated by an
analytical state coding and an evaluation of matrix elements of
the Heisenberg Hamiltonian ``on the fly".

To this end an analytical coding for the product basis states
\begin{eqnarray}
\label{magmol-E-4-5}
\ket{m_1, \dots, m_u, \dots, m_N}
\end{eqnarray}
is needed. Such a coding was already devised in
Ref.~\cite{SHS:JCP07} for equal spins. Here we show that this
scheme can be easily generalized for spin systems consisting of
different spins $s_i$, so that $-s_i\le m_i \le s_i$. 
For encoding purposes, and since $m_u$ can be
half-integer, the basis states are usually rewritten in terms of
quantum numbers $a_i=s_i-m_i$ instead of $m_i$, where
$a_i=0,1,\dots,2s_i$. 

The non-trivial technical problem of the coding stems from the
fact that one wants to use the $\op{S^z}$ symmetry, i.e. work in
subspaces ${\mathcal H}(M)$ of total magnetic quantum number
$M$. $M$ assumes values from $-\MMax$ up to $\MMax$ with
$\MMax=\sum_i s_i$. The basis in the subspace ${\mathcal H}(M)$ is given by all
product states $\ket{a_1, \dots, a_N}$ with $M=\MMax-\sum_i
a_i$. For usage in a computer program they need to be assigned
to integer numbers $1, \dots, \text{dim}\left({\mathcal
H}(M)\right)$. The reason is that one usually does not need the
basis only once at initialization, but at every Lanczos
iteration, since the sparse Hamiltonian matrix is not stored,
but its non-zero matrix elements are evaluated whenever needed
using
\begin{eqnarray}
\label{E-3-b}
\bra{i}\op{H}\ket{j}
&\equiv&
\bra{a_1^i, \dots, a_N^i}\op{H}\ket{a_1^j, \dots, a_N^j}
\ .
\end{eqnarray}
For a direct coding algorithm of basis states in
subspaces ${\mathcal H}(M)$ it is advantageous that the
sizes of the subspaces ${\mathcal H}(M)$ are known
analytically \cite{BSS:JMMM00}. Thus an array can be built at
startup that contains for a fixed sequence ${s_1, s_2, \dots,
  s_N}$ the sizes of these subspaces ${\mathcal H}(M=\MMax-A)$
for given a given number $n$ of spins and
$A$. We will call this array $\dimhm{N}{A}$. It will be used to
determine the sequential number of a basis vector in ${\mathcal
  H}(M)$.  The recursive buildup is performed using the
following relation between the sizes of subspaces
\begin{eqnarray}
\label{E-3-c}
\dimhm{n}{A}
&=&
\sum_{k=0}^{2s_n}
\,
\dimhm{n-1}{A-k}
\ .
\end{eqnarray}
If $A \notin \{0,1,\dots,2 \MMax\}$ then
$\dimhm{n}{A}=0$.\newline
For $\dimhm{n=1}{A=0,1,\dots,2s_n}=1$, $\dimhm{n}{A=0}=1$, and
$\dimhm{n}{A=1}=n$. If $A \notin \{0,1,\dots,2 \MMax\}$ then
\linebreak
$\dimhm{n}{A}=0$.

\subsection{$i \Rightarrow \ket{a_1^i, \dots, a_N^i}$}

One coding direction, $i \Rightarrow \ket{a_1^i, \dots, a_N^i}$,
which is the more trivial direction, can be realized in several
ways. A direct algorithm $i \Rightarrow \ket{a_1^i, \dots, a_N^i}$
using the known dimensions of the subspaces ${\mathcal
H}(M=Ns-A)$ could be realized as follows\footnote{The given code
uses FORTRAN notation. Nevertheless, it can be easily
transformed into C. One should only pay attention to the fact
that field indices in FORTRAN start at 1 not at 0. Therefore,
the definition of the second field index of $D$ has been
modified accordingly.}
\begin{verbatim}
      m=0
      Ak = A
      do k=N,2,-1
         do n=0,2*s(k)
	    if(i.le.(m+D(k-1,Ak-n+1))) then
	       BasisVector(k) = n
	       Ak = Ak - n
	       goto 100
            else
	       m = m + D(k-1,Ak-n+1)
            endif
         enddo
100      continue
      enddo
      BasisVector(1) = Ak
\end{verbatim}
\verb§BasisVector§ contains the $N$ entries $a_k$. This
algorithm will be made clearer when we explain the inverse
algorithm below.

\subsection{$\ket{a_1^i, \dots, a_N^i} \Rightarrow i$}

The inverse direction is actually the nontrivial one, since the
basis vectors are only a subset of the full basis set
\fmref{magmol-E-4-5}. Therefore, for the latter coding direction search
algorithms are often employed,\cite{PhysRevB.34.1677} or the
two-dimensional representation of Lin is
used \cite{PhysRevB.42.6561}.

The position of a basis vector $\ket{a_1, \dots, a_N}$ in the
lexicographically ordered list of vectors will be determined by
evaluating how many vectors lay before this vector. For this
purpose the known dimensions of the subspaces ${\mathcal
H}(M=\MMax-A)$ are used again.  In a computer program one can
evaluate the position $i$ of 
$\ket{a_1, \dots, a_N}$ in the list of basis vectors according
to
\begin{verbatim}
      Ak = A
      i = 1
      do k=N,2,-1
         do n=0,BasisVector(k)-1
            i = i + D(k-1,Ak-n+1)
         enddo
         Ak = Ak - BasisVector(k)
      enddo
\end{verbatim}
\verb§BasisVector§ contains the $N$ entries $a_k$.  If the
array of dimension $\dimhm{N}{A}$ is properly initialized,
i.e. the field value is zero for non-valid combinations of $N$
and $A$, then the sum can be performed in a computer program
without paying attention to the restrictions for the indices.



\end{document}